\documentclass[pre,aps,twocolumn,superscriptaddress]{revtex4-1}
\usepackage{graphicx}
\usepackage{url}
\usepackage{amsmath,amssymb}
\usepackage[version=3]{mhchem}
\usepackage{xcolor}

\begin{document}
\date{\today}
\title{Out of equilibrium stationary states, percolation,\\and sub-critical instabilities in a fully non conservative system}

\author{Mathieu G\'enois} \email{Mathieu.Genois@cpt.univ-mrs.fr}
\affiliation{Laboratoire Mati\`ere et Syst\`emes Complexes (MSC),
  Univ. Paris-Diderot, CNRS UMR 7057, F-75205 Paris CEDEX 13, France}
\affiliation{CPT, UMR 7332, Univ. Aix-Marseille} \affiliation{Academy
  of Bradylogists} 
\author{Pascal Hersen} \affiliation{Laboratoire
  Mati\`ere et Syst\`emes Complexes (MSC), Univ. Paris-Diderot, CNRS
  UMR 7057, F-75205 Paris CEDEX 13, France} \author{Eric Bertin}
\affiliation{LIPHY, Universit\'e Grenoble Alpes and CNRS, F-38000
  Grenoble, France} \author{Sylvain Courrech du Pont}
\affiliation{Laboratoire Mati\`ere et Syst\`emes Complexes (MSC),
  Univ. Paris-Diderot, CNRS UMR 7057, F-75205 Paris CEDEX 13, France}
\author{Guillaume Gr\'egoire} \email{guillaume.gregoire@ec-nantes.fr}
\affiliation{Laboratoire Mati\`ere et Syst\`emes Complexes (MSC),
  Univ. Paris-Diderot, CNRS UMR 7057, F-75205 Paris CEDEX 13, France}
\affiliation{HPC Institute (ICI), \'Ecole
  Centrale, Nantes, 1 rue de la No\"e, F-44300 Nantes, France}
\affiliation{Academy of Bradylogists}

\begin{abstract}
  The exploration of the phase diagram of a minimal model for barchan
  fields leads to the description of three distinct phases for the
  system: stationary, percolable and unstable. In the stationary phase
  the system always reaches an out of equilibrium, fluctuating,
  stationary state, independent of its initial conditions. This state
  has a large and continuous range of dynamics, from dilute -- where
  dunes do not interact -- to dense, where the system exhibits both
  spatial structuring and collective behavior leading to the selection
  of a particular size for the dunes. In the percolable phase, the
  system presents a percolation threshold when the initial density
  increases. This percolation is unusual, as it happens on a
  continuous space for moving, interacting, finite lifetime dunes. For
  extreme parameters, the system exhibits a sub-critical instability,
  where some of the dunes in the field grow without bound. We discuss
  the nature of the asymptotic states and their relations to well-known
  models of statistical physics.
\end{abstract}

\maketitle

\section{Introduction}
One of the key assumptions of equilibrium statistical physics is the
existence of conservation laws associated to quantities like energy,
linear and angular momenta, and number of particles. Out of
equilibrium, some of the conservation laws may break. In driven
systems, some of the mechanical quantities are continuously injected
and dissipated into the surrounding
medium~\cite{Aumaitre_2001_EPJB}. In reaction-diffusion
problems~\cite{Schlogel_1972_ZfP, Brower_1978_PLB}, even the
conservation of the number of particles may be absent. Systems without
conservation laws often exhibit an absorbing phase transition (APT)
between an active phase with a fluctuating number of particles, and an
absorbing phase without any activity. Depending on the model, it could
be a state where all particles have disappeared, or where particles
are in a frozen state. A prominent universality class for absorbing
phase transitions is the Directed Percolation (DP)
class~\cite{Grassberger_1982_ZPB, Hinrichsen_2000_AiP}. But the
definition of a class of universality is very sensitive to the
underlying symmetries: parity of the number of
reactants~\cite{Kockelkoren_2003_PRL}, nature of the absorbing
phase~\cite{Rossi_2000_PRL}, \emph{etc.}, imply different universality
classes from DP. Furthermore, if a source of noise has an effect on
the absorbing phase, it seems that the phase transition
disappears~\cite{Grassberger_1982_ZPB, Prakash_1997_JSP}.

In reaction-diffusion models, the dynamics is defined in terms of
particles, and the order parameter is linked to the number of
particles. In other models, the dynamics acts on an continuous
additive quantity: mass, energy, or momentum. Related phase
transitions happen between a low, even fluctuating, homogeneous level
of this quantity and a localized states where it is maximized. For
instance, in systems of self-propelled
particles~\cite{Vicsek_1995_PRL}, the momentum is zero on average in
the disordered phase, whereas it is concentrated in solitons, or in
non-linear, periodic peaks~\cite{Gregoire_2004_PRL, Solon_2015_PRL_a}
in the ``ordered'' phase. In mass transfer models
(MTM)~\cite{Evans_2005_JPAMT}, an out-of-equilibrium activity
maintains the exchange of mass between sites and leads to a transition
of mass condensation on few sites.

Previously~\cite{Genois_2013_GRL, Genois_2013_EPJB}, geophysical
matters have lead us to build a model for barchan fields in order to
understand the peculiar characteristics of such structures. Our
studies were based on experiments~\cite{Hersen_2004_EPJB,
  Hersen_2005_GRL} and on field observations~\cite{Bagnold_1941_book,
  Finkel_1959_JG, Elbelrhiti_2008_JGR}. However, we will now ignore
the natural background of the model to study its whole phase
diagram. We consider objects, which we will arbitrary call
\emph{dunes} and which are characterized by an extensive quantity $V$
that we will call \emph{volume}, but could as likely be either mass or
energy.  Those objects appear, move spatially, react with each other
and disappear depending on the value of $V$.

This model presents features similar to both reaction-diffusion models
and MTM. One can wonder whether a symmetry will govern the properties
of the system and its phase transition, or if we get a richer phase
diagram. We propose now to investigate the parameter space of our
system to understand the interplay between both ingredients.

In the following, we define our model. We question its microscopic
symmetries and we present the (classical) models of statistical
physics to which we expect to compare our dunes model. Then, with
numerical measurements and analytical arguments, we will show that
percolated noisy deserts can be found. Finally, in an opposite limit
of the control parameters, we find a transition of dune condensation.

\begin{figure}[t]
  \includegraphics[width=0.45\textwidth, clip]{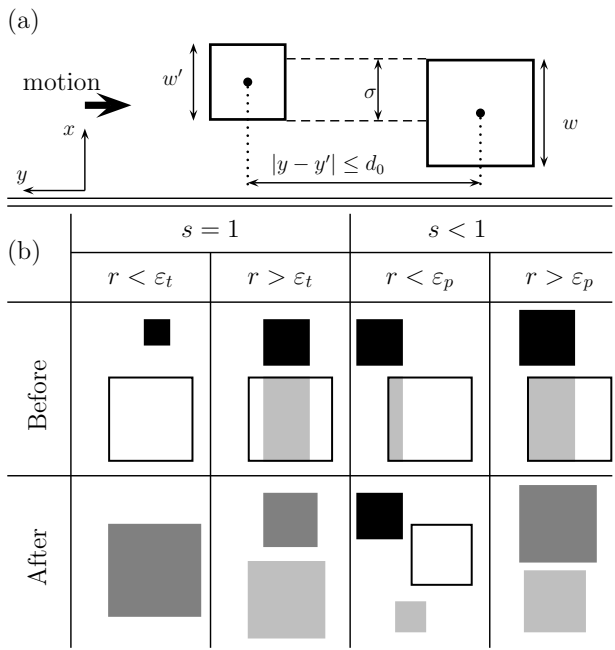}
  \caption{Interactions between dunes. (a) Remote interaction and
    definition of the neighborhood. (b) The four types of ideal collisions:
    merging, exchange with $s=1$, fragmentation, exchange with $s<1$.}
  \label{fig:inter}
\end{figure}

\section{Description of the model}
In this model, cubic dunes are labeled by their position $(x,y)$
on the field and their size $w = V^{1/3}$. These variables are continuous, as
neither the space nor the size are discretised. Dunes move on the field in the
decreasing $y$ direction only (Fig.~\ref{fig:inter}(a)), with a speed
$v$ inversely proportional to their size:
\begin{equation}
  v = \frac{\alpha}{w},\label{eq:kinematics}
\end{equation}
where $\alpha$ is a parameter which is related to the mobility of
  a dune.

The field has a length $L$ and a width $\ell$. We impose bi-periodic
conditions at the borders of the field, ensuring that dunes reaching
the $y=0$ limit (respectively $x=0$) are going on their ways at the
$y=L$ border (resp. $x=\ell$), and vice-versa.

The size of a single dune decreases in time according to the
following law:
\begin{equation}
  w(t) = \left(w^3(t_0) -(t-t_0)\times\Phi\right)^{1/3}\label{eq:loss}
\end{equation}
where $w(t_0)$ is the initial size of the dune, $t_0$ the time it
appears on the field, and $\Phi$ the constant rate of volume
loss. This law is valid until the size of the dune reaches the minimum
value $w_m$, when it is then removed from the field. To compensate for
this outflux, dunes of size $w_0$ are injected on the field, randomly
in time and space, at a constant mean rate $\lambda$ by unit of time
and surface.

Dunes interact with each other in two different ways. If two dunes are
closer than a distance $d_0$ along the $y$-direction and the overlap
length $\sigma$ between them, along the $x$-direction, is not zero
(see Fig.~\ref{fig:inter}(a)), the downstream one catches a part
$Q_\Phi$ of the volume lost by the upstream one, proportional to the
ratio $s$ between the overlap length $\sigma$ and the upstream dune
size $w'$:
\begin{eqnarray}
  s &=& \frac{\sigma}{w'} \label{eq:s}\\
  Q_\Phi &=& s\Phi \label{eq:sand_collect}
\end{eqnarray}
This defines an effective remote interaction of range $d_0$ between
dunes. Dunes exchange continuous amounts of volume. If there are
several dunes upstream within the distance $d_0$, the total catched
volume is a sum over all these neighbors, taking into
account the screening effect of a dune before one another.

As the speed of a dune is inversely proportional to its size, small
dunes travel faster than big ones, and therefore dunes can collide. A
collision occurs when two dunes overlap along the $x$-direction, and
the center of mass of the upstream dune passes the center of mass of
the downstream one in the $y$-direction. We emphasize that dunes of
our model are cubic, so their physical extents allow collisions on a
continuous space. However time is discretized. Therefore we need to
test whether a collision happens during a time step $\Delta t$ or not.
A collision is defined by the fact that the ordinates of two dunes
will be equal within $\Delta t$. We fix the time step at $\Delta t=1$.
The mobility $\alpha$ (see Eq.~\ref{eq:kinematics}) is used to tune
the rate of the dynamics. In the following, the system is studied for
a given maximum time~: $10^5\Delta t$.

We impose that the total volume engaged during a collision is
conserved, and that a collision only modifies the volume repartition
between dunes. Collision phenomenology is entirely determined by its
local geometry. The overlap between the two dunes defines sections of
the downstream one. These sections are considered separately for the
resolution of the collision. The ordinates of the dunes after
collision are set to the ordinate of the previous downstream
dune. Their abscissas are calculated as the barycenters of the
sections they are made of, which can lead to some effective lateral
diffusion. We define four types of binary collision (see
Fig.~\ref{fig:inter}(b)), depending on the value of the parameters $s$
(defined in Eq.~\ref{eq:s}) and $r$ defined with the width $w$ of
downstream dune as follows:
\begin{equation}
  \label{eq:r}
  r = \frac{\sigma}{w}
\end{equation}

When the overlap is complete ($s=1$), we compare $r$ to a limit value
$\varepsilon_t$. If $r<\varepsilon_t$, the two dunes merge; if $r>\varepsilon_t$, the
collision rearranges the total volume between the two dunes. The
overlapped section of the downstream dune becomes independent, the
remaining sections are merged with the upstream dune. If the
overlap is not complete ($s<1$), we compare $r$ to another limit value
$\varepsilon_p$. If $r<\varepsilon_p$, the collision splits the downstream dune into
two dunes; if $r>\varepsilon_p$ the volume is rearranged between the two
dunes as the ($s = 1, r > \varepsilon_t$) case. The quantitative effect on the
volumes is summarized in Eq.~\ref{eq:coll}, where braces mark
individual dunes and brackets dune conformations.

\begin{equation}
  \label{eq:coll}
  \left[
    \begin{array}{c}
      \{w^3\}\\\{w'^3\}
    \end{array}
  \right]
  \rightarrow\left\{
  \begin{array}{cl}
    \left[
      \begin{array}{c}
        \{w^3+w'^3\}
      \end{array}
    \right]
    &
    s=1,r<\varepsilon_t\\
    \\
    \left[
      \begin{array}{c}
        \{\sigma w^2\}\\
        \{w^2(w-\sigma)\}\\
        \{w'^3\}\\
      \end{array}
    \right]
    &
    s<1,r<\varepsilon_p\\
    \\
    \left[
      \begin{array}{c}
        \{\sigma w^2\}\\
        \{w^2(w-\sigma)+w'^3\}\\
      \end{array}
    \right]
    &
    \begin{array}{l}
      s=1,r>\varepsilon_t\\
      s<1,r>\varepsilon_p\\
    \end{array}
  \end{array}
\right.
\end{equation}

Depending on the volume ratio and on the relative distance along the
$x$-direction, the interactions may smooth out the volume difference,
or increase it. They may shift the dunes away, or align them toward
the same axis~\cite{Genois_2013_EPJB}.

\begin{table}[t]
  \centering
  \begin{tabular}{|c|c|l|c|}
    \hline
    $w_m$&$L$&Minimum size&$0.01$\\
    \hline
    $w_0$&$L$&Injection size&$0.1$\\
    \hline
    $d_0$&$L$&Limit interaction distance&$1$\\
    \hline
    $\Phi$&$L^3T^{-1}$&Volume loss rate&---\\
    \hline
    $\lambda$&$L^{-2}T^{-1}$&Injection rate&---\\
    \hline
    $\alpha$&$L^2T^{-1}$&Dunes mobility&$0.001$\\
    \hline
    $\varepsilon_t$&$\varnothing$&Limit value for $r$ when $s=1$&$0.5$\\
    \hline
    $\varepsilon_p$&$\varnothing$&Limit value for $r$ when $s<1$&$0.5$\\
    \hline
  \end{tabular}
  \caption{Parameters of the model: symbols, physical
    dimensions, significance and reference values.}
  \label{tab:parameters}
\end{table}

Eight parameters control the phenomenology of the model (see
Table~\ref{tab:parameters}): three length scales, three time scales
and two dimensionless parameters. Thus, according to
Buckingham~\cite{Buckingham_1914_PR}, we can build four dimensionless,
independent control parameters. We first define two aspect ratios:
\begin{equation}
  \delta = \frac{w_m}{w_0}  \label{eq:delta}
\end{equation}
\begin{equation}
  \Delta = \frac{w_0}{d_0} \label{eq:Delta}
\end{equation}
We now explicit the three time scales of the system. According to
Equation~\ref{eq:loss}, the lifetime $\tau_d$ of a single dune is:
\begin{equation}
  \label{eq:tau_d}
  \tau_d = \frac{w_0^3 - w_m^3}{\Phi}
\end{equation}
The typical time $\tau_n$ between two dune nucleations on a typical
surface $d_0^2$ is:
\begin{equation}
  \label{eq:tau_n}
  \tau_n = \frac{1}{\lambda d_0^2}
\end{equation}
The typical collision time $\tau_c$ is defined as the time for the
quickiest dune to reach the slowest one within the interaction range
$d_0$, without considering any other phenomenology. If there is no
exchange of volume, the slowest dune is $w_0$ wide. Therefore,
$\tau_c$ is:
\begin{equation}
  \label{eq:tau_c}
  \tau_c = \frac{d_0}{\alpha}\left(\frac{1}{w_m} - \frac{1}{w_0}\right)^{-1}
\end{equation}
Then we can build two control parameters that compare these three times:
\begin{equation}
  \label{eq:xi}
  \xi = \frac{\tau_d}{\tau_n} = \frac{w_0^3-w_m^3}{\Phi}\lambda d_0^2
\end{equation}
\begin{equation}
  \label{eq:eta}
  \eta = \frac{\tau_d}{\tau_c} = \frac{w_0^3-w_m^3}{\Phi}\frac{\alpha}{d_0}\left(\frac{1}{w_m} - \frac{1}{w_0}\right)
\end{equation}

The first one compares the relative importance of injection and
dissipation in the system. For low $\xi$, the volume loss
predominates; for high $\xi$, the injection is the main drive of the
system. The second one compares isolated and collisional dynamics. For
low $\eta$, the dunes lifetime is low compared to the typical
collision time, therefore dunes hardly interact. For high $\eta$,
dunes experience lots of collisions before disappearing from the
field.

Dunes are made of a collection of sand grains under the drive of the
wind. And so their kinematics is really non trivial (see
Eq.~\ref{eq:kinematics}). There is no way to consider these objects as
isolated systems under classic conservation
laws~\cite{Genois_2013_EPJB}. Even during collisions where the volume
is locally conserved (Eqs.~\ref{eq:coll}), the effective kinetic
energy and momentum are neither conserved. Because of the minimal size
$w_m$, the dunes injection $\lambda$, and the merging and
fragmentation collision, the number of dunes is not conserved
either. Neither is the total volume, as the injection rate $\lambda$
is constant and not tuned to compensate the loss due to $\Phi$ and the
effect of the minimum size $w_m$. This system thus follows no
conservation law at the scale of the field. Therefore, no prediction
of its large scale dynamics or phase diagram can be made on the basis
of conservation law arguments, as often done in statistical
physics. In this study, we focus on the numerical exploration of its
$(\xi,\eta)$ phase diagram. All the other parameters are kept constant
(see Table~\ref{tab:parameters}). The length scale is thus defined by
$d_0$, and the time scale by $\tau_c$, through $\alpha$. We tune $\xi$
and $\eta$ by changing the loss rate $\Phi$ and the injection rate
$\lambda$.

\begin{figure}[t]
  \includegraphics[width=0.45\textwidth, clip]{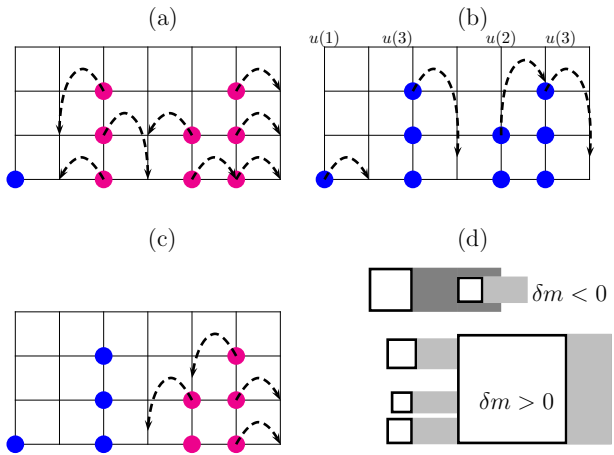}
  \caption{Definition of sand pile models. (a) BTW-Manna model: a pile of
    more than one particle is unstable. Then the grains are displaced
    (randomly) on neighboring sites. (b) Zero range process (ZRP): each
    grain moves at a rate $u$ which depends on the occupancy.  A
    variant, the misanthrop model, takes into account the occupancy of
  the departure and arrival sites. (c) Variant of BTW model where a
  site is activated by a neighbor. (d) Two configurations of remote
  sand exchange in our dune model. The dunes are moving from left to
  right. The shaded spaces figure out the sand which is collected by
the wind.}
  \label{fig:def_SP}
\end{figure}

\section{Analysis of symmetries\label{sec:analysis}}
One can first ask if there are some limits where the dynamics falls
  onto a well-known class of universality. When the dissipation is set
  to zero ($\Phi=0$), no more remote interaction occurs. The only
  events are the binary collisions, the nucleation and the disparition
  of dunes, named $A$ in the following. Annihilation happens because
  collisions can split dunes volume in a continuous manner, so the
  resulting volume can be less than the minimal volume $w_m$. The
  dynamics of Eq.~\ref{eq:coll} can be summarized as:
\begin{align}
\cee{2A&->[\alpha] A\label{eq:react_diff_barch_coal},\\
2A&->[\beta] 3A\label{eq:react_diff_barch_crea},\\
A&->[\mu] \varnothing\label{eq:react_diff_barch_anni},\\
 \varnothing&->[\gamma]A\label{eq:react_diff_barch_nuc}}.
\end{align}
The first three rules~(\ref{eq:react_diff_barch_coal},
\ref{eq:react_diff_barch_crea} and \ref{eq:react_diff_barch_anni})
embed this model in the pair-contact-process class. However, the
nucleation process (Eq.~\ref{eq:react_diff_barch_nuc}) makes the
absorbing phase fluctuate around a stationnary state. In Schl\"ogel
model~\cite{Grassberger_1982_ZPB, Prakash_1997_JSP}, such a noise is
known to smooth out the transition. We would like to understand how
the nucleation acts on our peculiar model.

One can also consider a \emph{quasi}-conservative limit where the
nucleation is set to zero ($\lambda=0$ in our dune model or $\gamma=0$
in Eq.~\ref{eq:react_diff_barch_nuc}). We call it
\emph{quasi}-conservative because we supress the source of sand, but
the persistence of dune annihilation still leads to a global decrease
of the total volume of sand. In that version, our model has a true
absorbing phase and is very close to the pair contact process with
diffusion (PCPD). The PCPD model has two
states~\cite{Kockelkoren_2003_PRL, Park_2005_PRL, Dornic_2005_arXiv,
  Hinrichsen_2006_PASMA}, one is an absorbing phase where at most one
particle diffuses. The second one is made of patches of persistent
activity. If the dynamics is figured on a spatio-temporal scheme,
those patches appear as percolated clusters along the time
direction. Since our dunes move in a ballistic way along the
$y$-direction, one can wonder whether dunes agregates percolate in
this direction.

Another way to analyse the rules of our model is to see each dune as a
sand pile (see Fig.~\ref{fig:def_SP} (a) and (c)). Without
dissipation, a single pile is stable and can to be destabilized by
another pile in its neighborhood. This results in a complex
reorganization of sand.  This is very similar to the
Bak-Tang-Wiesenfeld model~\cite{Bak_1987_PRL}, where piles of grains
are unstable above a threshold, but with a condition on the
neighbordhood (see for instance~\cite{Maslov_1996_PASMA}).

Such a comparison should come with many warnings. In particular, we
should discuss whether we are within the framework of self-organized
criticality (SOC)~\cite{Bak_1987_PRL} as it has been considered for
BTW model. It has been shown that SOC and APT are intrinsically
linked~\cite{Vespignani_1997_PRL, Vespignani_1998_PRL}. In SOC,
dissipation and driving are equal in magnitude, such that the global
density is constant, but their rates are decoupled. The APT
counterpart studies a model at a given density and its critical point
corresponds to the fixed point of the SOC model. This said, the way
the sand is dispatched in our model is deterministic. Deterministic or
random input~\cite{Manna_1991_JPAMT} is known to change the stationary
properties in a non-trivial way in sand pile
models~\cite{Ktitarev_2000_PRE, deMenech_2000_PRE}.

Although all of those points could act on the detailed dynamics, we
skip this discussion to concentrate on general aspects. The sand pile
model is known to exhibit agregates which go through the system in
avalanches or in multifractal waves, and its transition has common
features with critical phenomenon. Therefore we expect that our model
exhibits a transition to a phase where large agregates propagate along
the wind direction.

These considerations emphasize the role of sand exchange in contrast
with reaction process. We can wonder if there is another limit in
which reactions are no more the main process and are replaced as the
key-ingredient by the remote exchange of sand (see Eq.~\ref{eq:loss}
and Fig.~\ref{fig:inter}(a)). An obvious condition is to set the loss
of volume $\Phi$ to a high level. This sand is lost for ever if
there are no neighboring dunes, so the global density has also to be
sufficiently high to allow interactions. We will show that these
conditions are fulfilled at $\xi\gg 1$ and $\eta\le 1$. In this last
part of the phase diagram, the misanthrop model~\cite{Evans_2000_BJP}
can be a minimal model to understand our dynamics. In this model, a
variant of the zero-range process (see Fig.~\ref{fig:def_SP}(b)), an
element of mass goes from a site $i$ to the following one (in $d=1$)
with a rate $u(m_i,m_{i+1})$ which depends on the occupancy of both
sites~\cite{Evans_2005_JPAMT, Waclaw_2012_PRL,
  Evans_2004_JPAMG}. Depending on the rate of exchange, the spatial
distribution of mass in the system may exhibit a transition. In that
case, a small global density will remain homogeneously
distributed. But, above a critical density, the excess of mass
condense on a site. The condensate can move, and then the dynamics of
mass collection is explosive~\cite{Waclaw_2012_PRL}.

Although BTW-like models study unstable dynamics where sudden
rearrangements occur and spread like avalanches, the question in MTM
is rather to know what type of distribution is reached if sites retain
a part of the distributed mass. Indeed it is the main
difference. Other ingredients have been changed to test many variants
of the models. For instance, both models can be found with a bias
toward a direction~\cite{Hwa_1989_PRL, Dhar_1989_PRL}, or with
nucleation and sinks of matter~\cite{Evans_2005_JPAMT,
  Vespignani_1997_PRL, Vespignani_1998_PRL}. The mass can be
discretized, or can be a continuous variable~\cite{Basu_2012_PRL,
  Evans_2004_JPAMG}. Last, we have pictured these models on
one-dimensional space, but they also exist in any dimension. On that
peculiar point, even if the motions of barchans occur along the
$y$-direction, the interactions have true two-dimensional aspects as
it is depicted on Fig.~\ref{fig:inter}(b) and~\ref{fig:def_SP}(d).

\section{Out-of-equilibrium stationary states}

We first focus on the low $\xi$, low $\eta$ region. In this limit,
volume loss dominates both the dune injection and the collision
dynamics. As shown in~\cite{Genois_2013_GRL}, the system always
reaches a stationary, fluctuating, out-of-equilibrium state in which
dunes almost do not interact.
The dynamical properties remain normal, in the sense that macroscopic
quantities such as fluctuations of the number of dunes are
Gaussian~\cite{Genois_2013_EPJB}.

We then decrease the loss rate $\Phi$, which means that we travel
along the diagonal of the $(\xi,\eta)$ diagram, to the high $\xi$,
high $\eta$ limit. As $\Phi$ decreases, the density in the field
increases and interactions appear. Thus, the phenomenology changes,
and clusters of dunes appear in the field. They are created by the
destabilization of local high densities through avalanches of
collisions. We measured that the fragmenting collisions become
dominant in these structures and this generates lots of small dunes,
which can then catch up on other dunes. The dynamics of dune birth is
no longer Gaussian and this fact supports the idea of
avalanche~\cite{Genois_2013_EPJB}.

Inside a cluster, the density is high enough to prevent volume loss:
any volume lost by a single dune is catched by the downstream
ones. Definitive loss of volume happens mainly at the
downstream front of the cluster. Therefore, borders of these
structures are very well defined, as any dune put aside by a collision
loses volume and quickly vanishes.

Clusters are also responsible for a size selection in the field. As
they are very dense, dunes inside go through lots of collisions, whose
accumulation leads to the emergence of a new typical size
$\tilde{w}$. Whereas these dunes are small and would disappear quickly
in a diluted field, the effective conservation of volume in the
clusters stabilizes them. This selection is directly due to the
effective dynamics in the clusters and does not happen in the rest of
the field. It generates an anti-correlation between the local density
of dunes and the local mean width of dunes~\cite{Genois_2013_GRL}.

The crossover between dilute and dense dynamics is smooth, and
presents no sign that would mark the presence of a phase
transition. Quantities of the system evolve without any discontinuity
and their fluctuations do not diverge. Neither does the spatial
correlation along the $y$-direction~\cite{Genois_2013_EPJB}. This
crossover is merely a simple, smooth change of dynamics, due to the
progressive densification of the system. This can be seen as
counter-intuitive in the light of the analysis of symmetries
(section~\ref{sec:analysis}). One explanation is that the lifetime of
the agregate is never long enough to allow a clear breaking of
symmetry. To stabilize them, one can make the dynamics more
conservative in lowering $\Phi$ and $\lambda$. Another possibility is,
at a given dissipation $\Phi$, to increase the volume injection
$\lambda$. That is the subject of the next two sections.

\section{Percolation}
In the out-of-equilibrium stationary states, a very high activity
emerges within the agregates, although dunes barely interact in the
dilute regime. All these states are made of fluctuating populations of
objects as far as numbers and volumes of the objects are concerned.
However, the dilute regime has some features of an absorbing phase and
its fluctuations are given by the nucleation $\lambda$ and the loss of
sand $\Phi$. Suppose now that we suppress those two stochastic
processes, then the dilute regime will become a true absorbing phase:
without direct collision there is no way to produce or destroy
dunes. 

In presence of collisions however(see Fig.~\ref{fig:inter}), the number
of dunes may fluctuate whereas the total volume is kept constant. So
one can wonder whether it is possible to get enough collisions to
produce an active phase and a phase transition to this new state. In
the following, we address this question first in a quasi-conservative
system where $(\Phi,\lambda)=(0,0)$. Then we increase the level of
fluctuations of the volume $(\Phi,\lambda)\neq (0,0)$ to investigate the
robustness and the properties of the new phase.

\subsection{The quasi-conservative system}

Increasing $\eta$ as $\xi$ is kept constant is equivalent to
decreasing both the loss rate $\Phi$ and the injection rate $\lambda$
in the same manner, and thus lowering the non conservative aspect of
the system. We can even turn off the injection and dissipation. In
that case, $\xi$ is not defined anymore, and $\eta$ is
infinite. Notice that the existence of a minimal size $w_m$ maintains
a sink of matter.

When $\eta$ is sufficiently high, the behavior changes: the system
becomes sensitive to the initial conditions and exhibits a percolation
threshold when the initial density increases
(Fig.~\ref{fig:percol}(b)). This transition is rather unusual, as it
is a percolation of polydisperse, moving, interacting dunes on a
continuous space. Some systems with equivalent features have been
previously studied: continuous isotropic percolation of identical
disks~\cite{Grimmett_1999_book}, or squares and other anisotropic
objects~\cite{Baker_2002_PRE, Mertens_2012_PRE, Koza_2014_JSM}. Some
other models describe systems with an infinite number of degrees of
freedom~\cite{Corte_2008_NPhys}.

\subsubsection{Description}
\begin{figure}[t]
  \includegraphics[width=0.45\textwidth, clip]{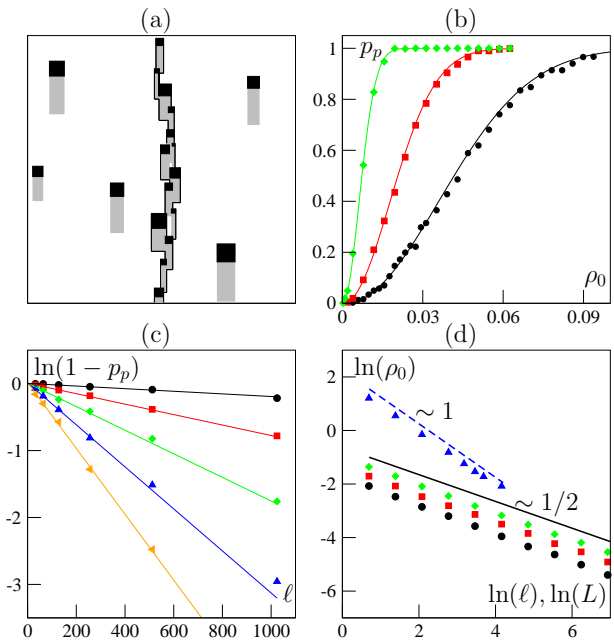}
     \caption{Percolation. (a): diagram of a percolation event on the
       field. Black areas are the proper surfaces of the dunes, gray
       areas are their interaction surfaces. The surface surrounded in
       black shows the cluster of dunes that percolates through the
       periodic boundaries. (b): probability $p_p$ for the system to
       percolate as a function of the initial density $\rho_0$, with
       $\Phi = 0$ and $\lambda = 0$, at a fixed length $L=16$, for
       different widths of the field : $\ell= 32$ ($\bullet$),
       128({\color{red}$\blacksquare$}), 1024
       ($\color{green}\blacklozenge$). The continuous lines are given
       by Eq.~\ref{eq:proba} without any fitting parameter. (c):
       $\ln(1-p_p)$ \emph{vs} $\ell$ for a fixed system length $L=16$
       and different initial densities ($\rho_0\simeq 0.0039
       (\bullet)$, $0.0078 (\color{red}{\blacksquare})$, $0.0117
       (\color{green}{\blacklozenge})$,
       $0.0156(\color{blue}{\blacktriangle})$,
       $0.0195(\color{orange}{\blacktriangleleft})$). The continuous
       lines are given by Eq.~\ref{eq:proba_scaling} without any fitting
       parameter. (d): finite size effects on the percolation
       transition. We plot the initial density $\rho_0$ needed to get
       a certain probability $p_p^0$ for the system to percolate : when
       $\ell$ varies and $p_p^0 = 0.25$ ($\bullet$), 0.5
       ($\color{red}\blacksquare$), 0.75
       ($\color{green}\blacklozenge$), $L=16$; when $L$ varies and
       $p_p^0=0.978$ and $\ell=2$ ($\color{blue}\blacktriangle$). The
       black continuous line shows a power law of exponent -1/2, the
       dashed (blue) one an exponent of 1. \label{fig:percol}}
\end{figure}

Each dune of size $w$ defines an interaction area of length $d_0$ and
width $w$ in front of itself. The surface of a dune is defined as the
reunion of the proper surface $w^2$ of the dune and its interaction
surface $wd_0$. We call percolation the onset of a path that connects
the upper and the lower border of the field through overlaps of
dunes surfaces, and whose extremities connect themselves through
the periodic boundaries (Fig.~\ref{fig:percol}(a)).

We compute the probability of percolation in counting the number of
percolated events for a given computation time. This probability $p_p$
evolves as the initial density $\rho_0$ is changed (see
Fig.\ref{fig:percol}(b)). We thus define numerically a threshold
$\rho$ when the probability reaches a given value $p_p^0$. 

Studying this probability allows us to check the existence of
percolated clusters with a low numerical effort. It is however not the
classical order parameter. To investigate the properties of the phase
transition, one has to study the probability for a dune to be inside
an infinite agregate.

Finite size effects show that when the width $\ell$ of the field
increases, the threshold tends to zero with a $\ell^{-1/2}$ law
(Fig.~\ref{fig:percol}(b) and (d)). If the length is increased, the
threshold vanishes as $L^{-1}$ (Fig.~\ref{fig:percol}(d)). We
emphasize that three very different values of $p_p^0$ have been used
to produce the figure~\ref{fig:percol}(d). So the whole curve is going
to be steeper and steeper as the width is increased. Therefore, we
could deduce that this percolation could appear at a zero density for
an infinite system. 

This result is quite atonishing at the light of former studies of
percolation on a continuous space~\cite{Grimmett_1999_book}, and for
different shapes of object~\cite{Baker_2002_PRE, Mertens_2012_PRE,
  Koza_2014_JSM}. Therefore, something in this analogy must be
misleading.  Indeed, the process which leads to a percolation event is
the fruit of dynamical interactions: there is no percolation without
collisions.

In the limit of a quasi-conservative system, we propose to assume
that a percolation event is the result of the interaction of two
dunes, which collide with each other and their daughters many times because
of the periodic boundary conditions, generating many new dunes in
their column and thus forming the percolating cluster.
Following this hypothesis, the system is then entirely
determined by its initial configuration, as there is no
nucleation. The probability for the system to percolate is therefore
simply the probability to find at least two dunes in the same column.

\subsubsection{Analytic arguments}
Considering that an aggregate is the consequence of an avalanche of
fragmenting collisions, let us consider how many interacting dunes are
needed to create a percolation event.  Initially, the system is fed
with a homogeneous distribution of dunes with a mean size $w_0$. For
the system to percolate, one must have at least two dunes of mean size
$w_0$ in the same column, colliding and then generating through
multiple collisions a minimal percolation structure, i.e., a column of
length $L$, of dunes of minimal size $w_m$, each separated from the
next downstream one by a length $d_0$.  So, to ensure that the two
initial dunes gather enough volume to generate the minimal percolation
structure, one has the mass balance:
\begin{equation}
\frac{L}{d_0}w_m^3\le 2w_0^3,\label{eq:balance}
\end{equation}
for a binary collision. Reversing this argument, we define here the
maximum size $L_2$ for a binary collision to create a percolation
event. For a longer system, one has to consider collisions with a
greater number of dunes. For a collision with a number $\mathcal{N}$
of similar dunes, the percolating cluster can reach a length of
\begin{equation}
L_{\mathcal{N}}=\mathcal{N}d_0\left(\frac{w_0}{w_m}\right)^3.\label{eq:min_size}
\end{equation}
There is no other role of the length in the quasi-conservative
system. Since dunes do not lose any sand when they are isolated, the
longitudinal distance delays the appearance of the percolation, but
does not prevent it in any other manner.

The width of the system might changes the probability of percolation
since the type of collision changes according to the relative lateral
position of dunes. For two similar dunes of width $w_0$, their
relative distance along $x$-direction has to be less than
$\varepsilon_p w_0$ (see Fig.~\ref{fig:inter} and Eqs.~\ref{eq:r}
and~\ref{eq:coll}). But collisions are symmetric along the axis of
motion. Therefore, the cross-section of this binary collision is
$d=2\varepsilon_p w_0$. Following our hypothesis, the percolation
probability $p_p$ might be the probability to find at least two dunes
of size $w_0$ in a column of width $d$. We compute the complementary
probability, namely the probability to find no more than one dune
in a column $p(n\le 1)$. We also assume that there are only
few dunes, so that the number of dunes $N = \rho_0L\ell$ is such as
$N\le\ell/d$.  We define $k=\ell/d$, the number of cross-section wide
columns within the field, then we find:
\begin{eqnarray}
p(n\le 1)&=&\frac{k(k-1)\ldots(k-N+1)}{k^N}\nonumber\\
&=&\frac{k!}{k^N(k-N)!}\label{eq:proba}
\end{eqnarray}
In Fig.~\ref{fig:percol}(b), we show the probabilities of percolation
$p_p$ obtained by changing $N$ while keeping all the other parameters
constant, for different widths $\ell$ of the system, and compare the
previous analytical result to these measures. We show that this simple
analytic argument models very well the data without the need of any
fitting parameter.
\begin{figure}[t]
  \includegraphics[width=0.45\textwidth, clip]{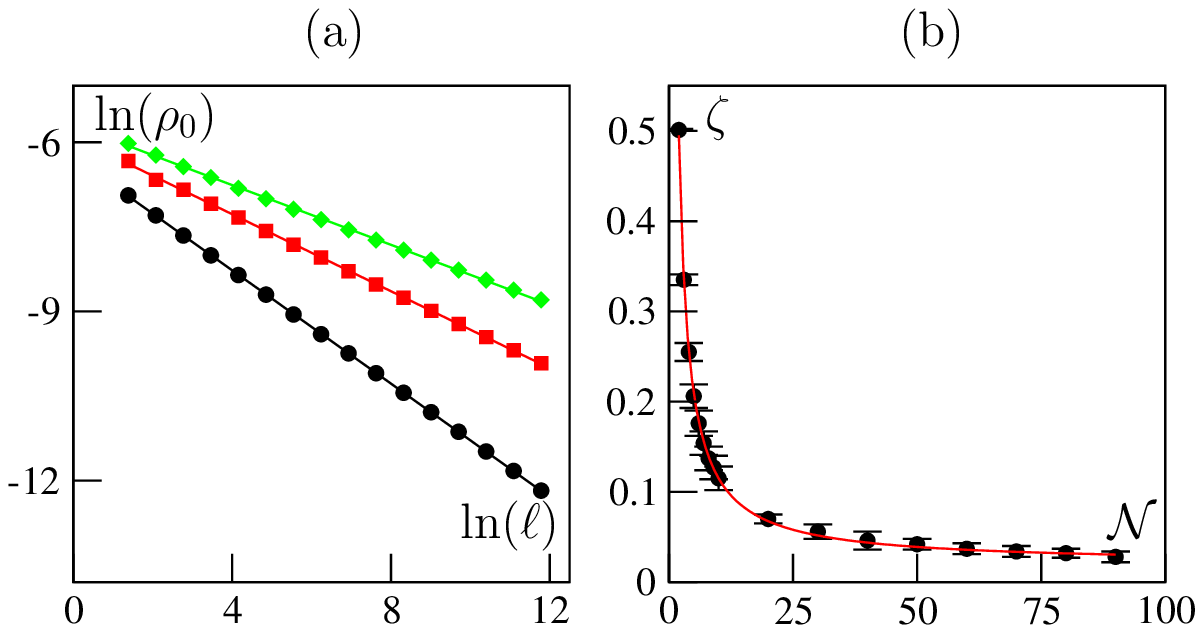}
\caption{Size effects in the probabilistic model. (a): the initial
  density $\rho_0$ which leads to a probability of percolation of
  $p_p=1/2$ \emph{versus} $\ell$ and for different $\mathcal{N}=2$
  ($\bullet$), 3 ($\color{red}\blacksquare$) and 4
  ($\color{green}\blacklozenge$). The plain lines are fitting curves
  which correspond respectively to an exponent $\zeta$ of $0.5$,
  $0.35$ and $0.28$ (in $\ln$--$\ln$ scale). (b): the exponent $\zeta$
  for the number $\mathcal{N}$ of colliding dunes. The plain line is
  the best algebraical fit~: $\zeta\simeq
  0.020+0.95/\mathcal{N}$.}\label{fig:fss_model_Ll}
\end{figure}
\subsubsection{Finite size effects}
We already have shown that the probability of percolation changes from
a system to another with different sizes
(Fig.~\ref{fig:percol}(b-d)). Since the percolation happens in the
axis of motion, and while there is some lateral diffusion, length and
width act differently on the value of $p_p$. So we look at the finite
size effects in decorrelating width and length.

Keeping the length constant, we define $\kappa=\rho L d$. Then,
remembering that $k=\ell/d$, we re-write the denominator of
Eq.~\ref{eq:proba}, $k-N=k(1-\kappa)$ which can be taken arbitrary
large for any $\kappa<1$. So we use Stirling's approximation in
Eq.~\ref{eq:proba} and we find the following scaling when
$k\to\infty$:
\begin{equation}
\ln\left(p(n\le 1)\right)\sim 
-\left[(1-\kappa)\ln\left(1-\kappa\right)+\kappa\right]k
-\frac12\ln\left(1-\kappa\right),\label{eq:proba_scaling}
\end{equation}
We indeed observe that the percolation probability, $p_p=1-p(n\le 1)$,
tends to one exponentially as the width is increased, see
Fig.~\ref{fig:percol}(c), following the exact scaling of
Eq.~\ref{eq:proba_scaling}. 

The constraint $\kappa<1$ is reminiscent of the fact that a greater
number of dunes than the number of columns obviously leads to a
percolated system. So, if we keep $\ell$ constant and increase the
length $L$ at a given density, the probability of percolation
increases to one where we expect $\kappa\sim 1$:
\begin{equation} 
p_p\sim 1 \Rightarrow \rho_0\propto \frac{1}{L}.
\end{equation}
We observe such a scaling on data, Fig.~\ref{fig:percol}(d), for
relatively small system sizes ($L\le 64$). Simulating larger systems
is just a matter of computation time. Let us also point out the fact
that our numerical systems were never long enough to test the mass
balance of a percolating cluster (Eq.~\ref{eq:min_size}). But one can
have an idea of the effect of $L$ using our probability model: when
$L\le L_2$, only binary collisions occur. If
$L\in[L_{\mathcal{N}-1};L_\mathcal{N}]$, one has to consider
collisions involving $\mathcal{N}$ dunes. 

In other words, we have to study the probability to find at least
$\mathcal{N}$ dunes in one column. We assume that the cross section
remains $2\varepsilon_p w_0$. To argue for this point, let us
decompose the interaction of three dunes into two collisions. In the
first one, if positions of dunes are homogeneous, the mean lateral
position is $\varepsilon_p w_0/2$, which leads to a new dune with a
volume $\varepsilon_p w_0^3/2$, or a width
$w_1=w_0(\varepsilon_p/2)^{1/3}$ (see Eq.~\ref{eq:coll}). The later
bumps into a dune of mean width $w_0$, for which the maximum
cross-section is $\varepsilon_p w_0$. We notice also that
$w_1\ge\varepsilon_p w_0$ as soon as $\varepsilon_p\le\sqrt{2}/2$. In
our simulations $\varepsilon_p=0.5$, therefore the cross-section is
the maximal cross-section. Considering the symmetry along the axis of
motion, we conclude that the discretization of the space in
$2\varepsilon_p w_0$-wide columns is still valid.

We computed the probability $p(n\ge \mathcal{N})$ to find at least one
column of width $d$ with at least $\mathcal{N}$ over a total number
$N$ of dunes at a given width $\ell$. For small lengths, we observe on
Fig.~\ref{fig:fss_model_Ll} that the percolation threshold, defined as
$p(n\ge \mathcal{N})=1/2$, vanishes as $\ell^{-\zeta}$. But the exponent
decreases with the number of dunes $\mathcal{N}$. The data are
consistent with a non-vanishing asymptotic value for $\zeta$, which
would mean that the transition occurs at any density for any width and
length in this mean-field model.

\begin{figure}[t]
  \includegraphics[width=0.45\textwidth, clip]{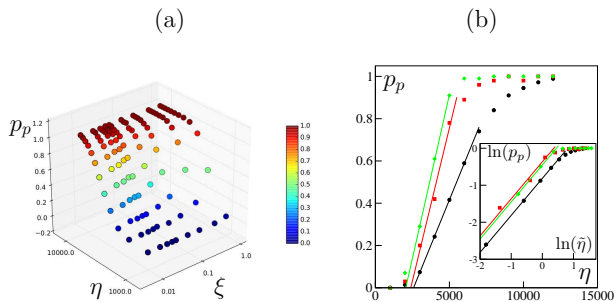}
  \caption{Percolation in non-conservative systems. (a): phase diagram
    for different sets of parameters $(\eta,\xi)$ at a fixed system
    width $\ell=32$ in log-log scale. The color map indicates the
    magnitude of percolation probability~: the probability decreases
    with increasing $\eta$. (b): transitions at $\xi=0.01$ for varying
    $\eta$ and for different system widths $\ell=32 \bullet$, $64
    \color{red}{\blacksquare}$, $128
    \color{green}{\blacklozenge}$. The continuous lines are linear
    fits of the data. In the insert, the same data in $\ln$-$\ln$
    curve, with $\tilde\eta=1\!-\!\eta/\eta_c$. Other parameters of
    simulation $\rho=1/8$ and $L=32$.} \label{fig:perco_nonc}
\end{figure}

\subsection{Percolation with fluctuations}

In the quasi-conservative system, the control parameter of the
percolation is the initial density of dunes $\rho_0$ in the field. The
randomness is due to the initial conditions, that then determine
entirely the evolution of the system. On the contrary, the stationary
phase at low values of $(\xi, \eta)$ is stochastic, independant of the
initial condition~\cite{Genois_2013_EPJB} and the stationary density
is set by the dynamics to $\rho=\xi/d_0^2$. So, we now question the
existence of a percolation phase in a stochastic non-conservative
system, even though there are only few events of nucleation per time
step.

Indeed, the system still percolates. We scanned the $(\xi,\eta)$
diagram and measured the probability of percolation $p_p$ at a given
initial density $\rho_0$, see Fig.~\ref{fig:perco_nonc}(a). We observe
a zone where the system never percolates, and a region where
percolation becomes likely. The probability to percolate seems to vary
linearly between both regions and become steeper as the system size is
increased. Finite size effects thus confirm the existence of two
regions: with or without percolated cluster even in the presence
of dissipation and nucleation (Fig.~\ref{fig:perco_nonc}(b)).

At rather small system sizes, the behavior of the system turns out to
depend on the initial density $\rho_0$. As the density increases, the
percolation is more and more likely (data not shown). But the
transition line in the $(\xi,\eta)$ diagram is not shifted by the
variation of the initial density. As we have shown that percolation
appears in quasi-conservative system even at very low density for very
large system, we may think that this dependency is a finite-size
effect. We observe that the percolation probability also increases
when the system size is increased. The number of nucleations increases
with the system size, it makes the system more stochastic, but it also
feeds the system with an amount of sand and makes the percolation
likely to appear. Hence, it allows the system to lose the memory
of its initial conditions.

\subsection{Percolation dynamics}
\begin{figure}[t]
  \includegraphics[width=0.45\textwidth, clip]{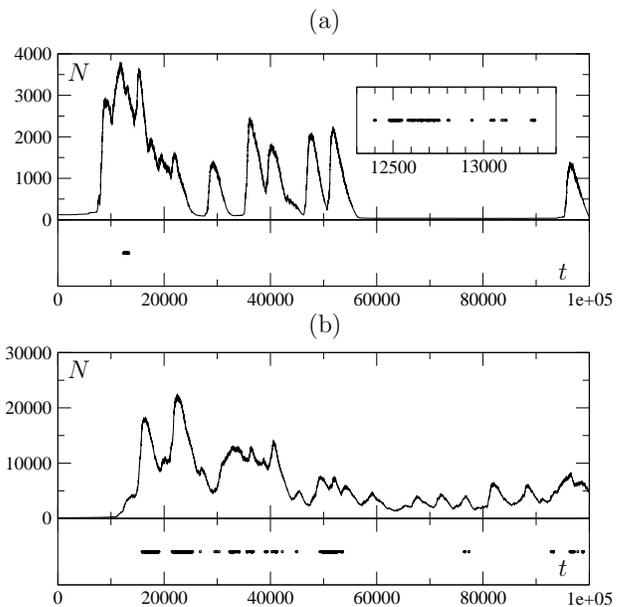}
  \caption{Modes of percolation. Evolution of the number of dunes in
    the field (top of each sub-figure) and events of percolation
    (bottom) at a given $\eta=10^4$ and for two different $\xi$ : $\xi
    = 0.01$ in (a) and $\xi = 1$ in (b). The insert shows a zoom of
    the only succession of percolation events that occurs in (a).
    Other parameters of simulation $\rho=1/8$ and $L=\ell=32$.}
 \label{fig:percol_mode}
\end{figure}
A percolation event is not a stationary pattern, even for a
quasi-conservative system. Fragmenting collisions split dunes,
whatever their sizes, into smaller objects. They can thus become smaller
than the smallest possible dune, and disappear. That is why an avalanche of
collisions that created a percolated cluster erodes it after a
while. For any finite $\eta$, volume loss occurs. However we expect it
has little effect on a percolated aggregate, since a cluster is a zone
where sand is almost a conserved quantity. The process which will
dominate the dynamics of percolation will thus be the nucleation.

Indeed, for high values of $\eta$, the system easily percolates for
any value of $\xi$. However, the temporal evolution of the system
differs a lot along the $\xi$ range. Percolation usually occurs the
first time during the transient regime. Then, the system can rebuild a
percolating situation through nucleation, and other events can
occur. This rebuilding takes a certain time, related to the nucleation
rate. Furthermore, not all clusters percolate, which reduces again the
probability for the system to present such an event. In
Fig.~\ref{fig:percol_mode}(a) we clearly see bursts of the number of
dunes, each signing the apparition of a cluster in the field, but only
one succession of percolation events. For low $\xi$, the time between
two series of percolation events is thus very large. For high $\xi$
this time tends to become rather small as the nucleation is more important
and rapidly refills the field. Indeed, in
Fig.~\ref{fig:percol_mode}(b) we have percolation series along the
whole simulation.

There are thus two asymptotic modes for the percolable phase, which
are closely related to the two modes of the stationary phase: a
dilute mode where percolation events are separated by very long
times, and a dense mode where percolation events occur much more often. As for
the stationary phase, a smooth crossover connects these two
modes. Moreover, in the $(\xi,\eta)$ space the two crossovers have
the same structure, connected above the phase transition line. This
points out that there might actually be only one dynamical crossover,
coming on top of the phase diagram (Fig.~\ref{fig:dp}).

\section{Giant dunes instability}

We now focus on the opposite limit, where $\eta$ is kept constant at a
low value and $\xi$ increases. In this limit, the volume loss is kept
at a high value by $\eta$, and the nucleation rate increases with
$\xi$. This is therefore a limit of high forcing and dissipation.

\begin{figure}[t]
  \includegraphics[width=0.45\textwidth, clip]{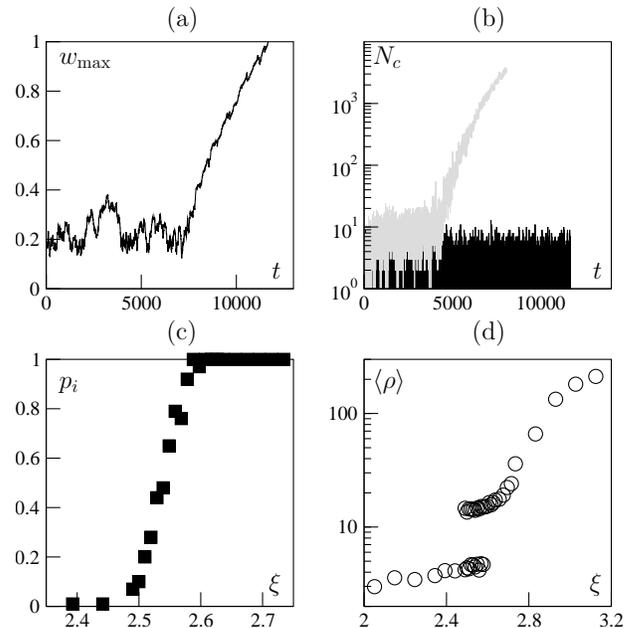}
  \caption{Giant dunes instability. (a): size of the biggest
    dune in the field, as a function of time, for $\xi = 2.15$ and
    $\eta = 0.3$. (b): collision rate as a function of time, for $\xi =
    2.15$ and $\eta = 0.3$ (black) and $\xi = 2.73$ and $\eta = 1$
    (gray). (c): probability for the system to be unstable as a function
    of $\xi$, for $\eta = 1$. (d): mean density $\langle\rho\rangle$
    at the end of the simulation as a function of $\xi$, for $\eta =
    1$, calculated on 100 realizations. Other parameters, see caption of
    Fig.~\ref{fig:percol_mode}.} 
  \label{fig:instab}
\end{figure}

When we increase the injection rate, the system is first homogeneous
and stationary. Then a critical $\xi$ appears beyond which the steady
state becomes unstable. After some time, the sizes of several dunes
begin to grow and never saturate (Fig.~\ref{fig:instab}(a)). If
the fixed value of $\eta$ is very low, this instability occurs in a
rather diluted field, with few collisions. For higher values of
$\eta$, a collisional stage occurs before the instability starts
(Fig.~\ref{fig:instab}(b)). As for both previous phases, this defines
two modes for the instability.

In the non-collisional mode, the system reaches a metastable state
before the instability starts (Fig.~\ref{fig:instab}(a)). For both
modes, there is no precise, critical value for $\xi$ but a range of
values where the probability for the system to develop the instability
grows continuously from 0 to 1
(Fig.~\ref{fig:instab}(c)). Furthermore, the mean values of
physical observables --for example the density-- measured at the
end of the simulation, present two disconnected branches, for the
stable and the unstable phase (Fig.~\ref{fig:instab}(d)). The
  observation of metastability and hysteresis is an indication of the
  fact that the instability is sub-critical.

The instability can appear in a low collisional system, therefore its
origin is probably not the merging type of collision. Indeed, a toy
model where a dune of size $w$ is randomly impacted by dunes of size
$w'$ shows that even for very large $w$, no instability involving only
collisions can appear. Even though the coalescence could in theory continuously
increase the size of a dune, the fragmentation is far more efficient at
decreasing this size (Fig.~\ref{fig:inter}).

The mechanism of the giant dunes instability rather involves the
remote interaction through volume exchange. The field at low $\eta$
and high $\xi$ contains a high number of dunes, which lifetimes are
very small due to the high volume loss. There are thus lots and very
important volume exchanges in the field. Every dune loses a volume
$\Phi$ per unit of time, but also gathers sand lost by any other
upstream dunes. The balance of sand strongly depends on the dune width
(see Fig.~\ref{fig:def_SP}(d)). For instance, let a dune of
size $w$ be followed by several dunes of size $w_m$ (which will
disappear next time step). The maximum sand balance will be when dunes
cover its whole size:
\begin{equation}
\frac{dV}{dt} = -\Phi+\Phi\frac{w}{w_m}.
\end{equation}
The small dunes will feed the large one and disappear without having
the time to collide with it.  So any local fluctuation of
dunes density behind a larger dune will make the latter grow, thus
increase its lifetime, and its ability to collect more sand. If the
injection rate is high enough, these collecting events are numerous
enough to make the size of some particles diverge, and generate the
instability.

The crossover from non-collisional to collisional instability is
smooth. As for the percolable phase, it is in fact the same crossover
that exists between dilute and dense stationary states.

\section{Conclusion}
\subsection{Summary}
We explored in this paper the phase diagram of a non trivial system.
The effective energy and momentum, as well as the number of dunes and
the total volume of these dunes are not conserved at the scale of the
whole system. The model has a peculiar phenomenology, inspired by the
geophysical problem of barchan fields. In particular, dunes interact
with each other through non trivial collisions and remote volume
exchange. Dunes are injected in the field, while volume loss at each
one of them ensures they have a finite lifetime. Two parameters,
comparing forcing and dissipation for $\xi$, isolated and interacting
behavior for $\eta$, define the phase diagram.

For standard values of the parameters, the system always reaches a
stationary state. Its dynamics range smoothly from non interacting to
interacting as both parameters increase, and are independent of the
initial conditions.

When $\eta$ becomes large, i.e., when the dissipation decreases, the
system becomes percolable, meaning that depending on the initial
density the system can exhibit a percolation transition.  This
percolation is unusual, as it occurs on a continuous space with
polydisperse, moving, finite lifetime, interacting objects. Indeed, we
show that for a system with an infinite width, percolated agregates
are likely to appear for any small value of density. An analytic,
mean-field, probabilistic model reproduces well the behaviour of the
probability to percolate. We extend the study of this model on the
effects of the system length, and it gives clues to suppose that
percolation is robust also when the length is increased. Similarly to
the stationary phase, dynamics range from dilute, where percolation
events are sparse in time, to dense, where they occur much more often.

When $\xi$ becomes large for low $\eta$, i.e., when both dissipation
and forcing are large, the system becomes unstable. Trapped in local
high densities, the sizes of some of the dunes grow without limit. The
instability, characterized by a discontinuity in the evolution of the
system observables, by a range of coexistence between stable an
unstable phase, and by a metastability before the beginning of
the instability, is sub-critical. As for the previous phases, its
dynamics smoothly range from non-collisional, where the collision rate
is low before the instability, to collisional, where the instability
begins after a large increase of the collision rate.

\begin{figure}[t]
  \includegraphics[width=0.45\textwidth, clip]{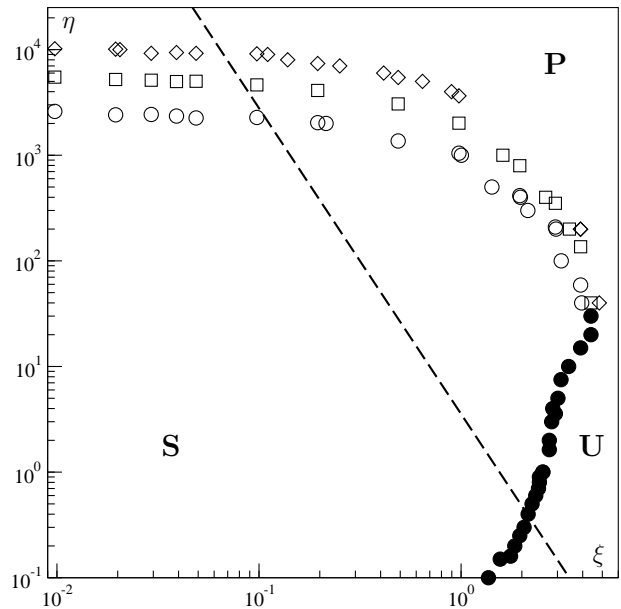}
  \caption{Phase diagram of the system. \textbf{S} is the stationary
    phase, \textbf{P} the percolable one, \textbf{U} the unstable
    one. The frontier between \textbf{S} and \textbf{P} is marked by
    three isolines for the probability for the system to percolate
    with $\rho_0 = 1$: $p_p = 0.05$ ($\circ$), $p_p = 0.5$
    ($\square$), $p_p = 0.95$ ($\lozenge$). The frontier between
    \textbf{S} and \textbf{U} is marked by the isoline for the
    probability for the system to be unstable $p_i = 0.5$
    ($\bullet$). The dashed line marks the smooth crossover from non
    collisional to collisional dynamics. Other parameters, same as
    Fig.~\ref{fig:percol_mode}.}
  \label{fig:dp}
\end{figure}

A smooth phase transition separates the stationary and the percolable
phase, a coexistence range separates the stationary and the unstable
phase, whereas the limit between percolable and unstable remains
unknown. In the end, the phase diagram of the system seems to consist
in the three previous proper phases, plus a dynamical diagram on top
of it. Indeed, the smooth range of dynamics from non-collisional to
collisional is found on all three phases, and is connected through
their limits (Fig.~\ref{fig:dp}). The parameters $\xi$ and $\eta$
thus define both the phase of the system, and the dynamics this phase
is exhibiting.

\subsection{Analogies and  future work}
Changing the relative values of $\xi$ and $\eta$ changes the relative
weight of the exchange of volume in the remote interaction
(Eqs.~\ref{eq:loss}, \ref{eq:s} and \ref{eq:sand_collect}) compared to
the local collisions (Eq.~\ref{eq:coll}). When collisions dominate, we
indeed found percolated cluters as in absorbing phase transition
models. When eolian transfer of mass are more frequent, mass
condensation occurs as in mass transfer model.

The first new point is the connection between both regions, a domain
of the phase diagram whose properties deserve to be studied. Next,
percolation seems to survive to fluctuations in contrast to classical
results on Schl\"ogel model. However this result is not so much
questioned. First fluctuations remain at a low level ($\xi \le 1$)
when percolation is likely. Second, when fluctuations increase in
comparison to dissipation, there is a smooth cross-over between
non-collisional and collisional dynamics and this cross-over is not a
phase transition. A last interesting test would be to modify the
nucleation process in order to have a true absorbing phase: for
instance one can nucleate a new dune close to a previously existing
dune.

Obviously, one has to characterize the phase transition of percolation
in our model. In eq.~\ref{eq:react_diff_barch_coal}
and~\ref{eq:react_diff_barch_crea}, rates of reaction depend on $\eta$
and $\xi$ but also on thresholds $\varepsilon _p$ and $\varepsilon
_t$. Therefore we could expect that the transition of percolation
should depend on them. But it is surprising that we do not need
$\varepsilon _t$ in our mean-field approximation.

The last region of the phase diagram is reached when the remote
exchange of sand dominates. We then observe a first order phase
transition of condensation. Mass transport models exhibit also a phase
with a condensate. Their stationary solutions are usually made of two
asymptotic phases: one is a nearly homogeneous density, the second
is made of a condensation of the excess mass to the latter homogeneous
repartition~\cite{Evans_2000_BJP}. In some case the dynamics of
condensation can be explosive~\cite{Waclaw_2012_PRL} and the
condensate visits ballistically the system. However, it seems to us
that the existence of a metastable state is a new feature of this
class of model. The explanation of this difference has probably to be
found in the non-conservative properties of our model.

\end{document}